\documentclass[conference]{IEEEtran}
\IEEEoverridecommandlockouts
\usepackage{amsmath,amssymb,amsfonts}
\usepackage{algorithmic}
\usepackage{graphicx}
\usepackage{textcomp}
\usepackage{subcaption,xcolor}
\usepackage[bookmarks=false]{hyperref}
\usepackage[numbers]{natbib}

\usepackage{textcomp}

\usepackage{makecell}

\def\BibTeX{{\rm B\kern-.05em{\sc i\kern-.025em b}\kern-.08em
    T\kern-.1667em\lower.7ex\hbox{E}\kern-.125emX}}
\begin{document}

\title{Detecting Compromised Implicit Association Test Results Using Supervised Learning
    \thanks{
    \textcopyright{} 2018 IEEE.  Personal use of this material is permitted.  Permission from IEEE must be obtained for all other uses, in any current or future media, including reprinting/republishing this material for advertising or promotional purposes, creating new collective works, for resale or redistribution to servers or lists, or reuse of any copyrighted component of this work in other works.
    Special thanks to P. Assumpcao, F. Bassey, E. Da Silva, T. Donnelly, I. Gomes, Y. Machado, K. Malone-Miller, L. Silva, and C. Snyder for their early contributions to this project. Research funded by the 2016 NSF REU program through Siena College.} 
}

    \author{
    \IEEEauthorblockN{Brendon Boldt}
    \IEEEauthorblockA{
    \textit{Marist College}\\
    Poughkeepsie, NY, USA \\
    brendon.boldt@gmail.com}
    \and
    \IEEEauthorblockN{Zack While}
    \IEEEauthorblockA{
    \textit{University of Massachusetts Amherst}\\
    Amherst, MA, USA \\
    zwhile@cs.umass.edu}
    \and
    \IEEEauthorblockN{Eric Breimer}
    \IEEEauthorblockA{
    \textit{Siena College}\\
    Loudonville, NY, USA \\
    ebreimer@siena.edu}
    }

\maketitle

\begin{abstract}
An implicit association test is a human psychological test used to measure subconscious associations.  While widely recognized by psychologists as an effective tool in measuring attitudes and biases, the validity of the results can be compromised if a subject does not follow the instructions or attempts to manipulate the outcome.  Compared to previous work, we collect training data using a more generalized methodology. We train a variety of different classifiers to identify a participant's first attempt versus a second possibly compromised attempt. To compromise the second attempt, participants are shown their score and are instructed to change it using one of five randomly selected deception methods.
Compared to previous work, our methodology demonstrates a more robust and practical framework for accurately identifying a wide variety of deception techniques applicable to the IAT.
\end{abstract}


\section{Background and Motivation}

\subsection{Implicit Association Test}

An implicit association test (IAT) is a human cognitive test that measures subconscious association between concepts and attributes \cite{Greenwald1998}. For example, in our experiment, the concepts are \textit{computer science} are \textit{biology}, and the attributes are \textit{male} and \textit{female}.  In our example, a positive IAT score indicates that a participant associates \textit{computer science} with \textit{male} and \textit{biology} with \textit{female}, whereas a negative scores indicates the participant associates \textit{computer science} with \textit{female} and \textit{biology} with \textit{male}.  Scores close to zero indicate weak or neutral associations.

Participants are shown items (typically words or images) on the center of a web page.  Each item represents one of the concepts or attributes. For example, in our experiment, first names such as \textit{James} and \textit{Mary} are used to represent the attributes \textit{male} and \textit{female} and words such as \textit{Internet} and \textit{habitat} are used to represent the concepts \textit{computer science} and \textit{biology.}

An IAT is divided into practice blocks, which familiarize the participant with the items, and critical blocks, which test the participant's ability to correctly and quickly categorize the items.  In each block, concepts, attributes, or both are assigned to the left and right side in different configurations (see  \cite{Greenwald1995}). When a participant is shown an item, they must press a keyboard key with either their left or right index finger to match the item with the correct concept or attribute. 

The key principle of an IAT is that if a participant naturally associates a concept with an attribute, they will be able to match items more quickly and with fewer errors when the concept and attribute are paired on the same side. Conversely, if a participant disassociates a concept with an attribute, their response will typically be slower and more prone to error when the concept and attribute are paired on the same side. The difference in response time and error rate between concept-attribute pairing is recognized by psychologists as a good measure of implicit or subconscious association. Implicit association is often different than self-reported association and detecting this difference is considered valuable to understanding biases, attitudes, and mental processes. \cite{Greenwald2005,Greenwald2009,Hofmann2005,Bargh2013}
By one survey, IATs accounted for $50\%$ of implicit bias measurements in social cognition research and since the time of the survey has remained an influential measurement tool. \cite{Nosek2011, Dimensions}

\subsection{Faking, Deceiving, and Compromising IAT Results}

The validity of IAT results can be compromised in many different ways.  Participants may not correctly follow the instructions or may become distracted during critical blocks. Participants can deliberately make errors or delay/accelerate their response time in order to alter their IAT score.  Participants can modify their response patterns through concentration or by physical modification such as changing their hand position. It is possible for the same participant to produce test results that show significantly different or contradictory outcomes.  For example, a participant's first IAT attempt could show a strong association between \textit{computer science} and \textit{male} but a subsequent attempt could show a very weak association or the opposite association, i.e.,  strong association between \textit{computer science} and \textit{female}. One of the goals of tools measuring implicit (subconscious) bias like an IAT is to rule out conscious or controlled responding. A critical problem is that participants who are familiar with IATs can potentially control results and deceptively obtain that does not reflect their natural implicit association.

Previous psychological studies have developed simple measures for identifying participants who faked their results.
By examining the difference between the average response time of the fastest critical block in a natural IAT and average response time of the slowest critical block in a faked IAT, researchers \cite{Cvencek2010}  averaged $75\%$ accuracy in identifying faked attempts.
By examining the ratio between average response time for the fastest critical block and the corresponding practice blocks, researchers \cite{Agosta2010}  achieved $80\%$
accuracy.  While these indices accurately identify scores that were faked by slowing down, more recent work \cite{Rohner2013} found that these indices fail to identify other deception strategies.

While many IAT faking strategies have been studied \cite{Steffens2004,McDaniel2009,Fiedler2010}, research in detecting IAT faking \cite{Stieger2011,Agosta2010,Cvencek2010,Rohner2015}  focus on only one or two deception strategies.   To obtain training data for analysis, researchers (i) use surveys to prune out participants who are familiar with IATs, (ii) use direct observation to prune out participants who are not following the instructions correctly,  and (iii) use aggregate data and cutoffs to prune out participants who cannot significantly alter their IAT score when instructed to do so. Thus, researchers are analyzing data consisting of only verified natural attempts and successfully faked attempts, which will have significant statistical differences. Thus, these studies give practitioners false confidence that determining the validity of IAT results is a simple and easily solved problem.

When one considers scenarios with subjects who cannot be surveyed or observed, who may not be following the correct instructions, or who may be employing a wide variety of effective or ineffective deception strategies, the problem of determining IAT validity is more challenging.  In this paper we focus on experiments that do not necessarily rely on the the aid of direct observation, prior knowledge about the participants, or selective data pruning.

\section{Methodology}

\subsection{IAT Implementation}

While IAT data is publicly available, it was important to obtain very specific training data to capture a variety of different deception techniques. Thus, we implemented our own online IAT to match the block configuration and improved scoring algorithm described in \citet{Greenwald2003}. 

{\footnotesize
\begin{table}[!t]
\centering
 \begin{tabular}{|l|l|} 
 \hline
\textbf{Concept/Attribute} & \textbf{Items}\\
\hline
Computer Science &	\thead{Apps, Computer, Algorithm, Database,\\Internet, Programming, Software, Technology} \\ 
 \hline
Biology & \thead{Nature, Life, Photosynthesis, Habitat, Organs,\\Plants, Species, Protein} \\ 
 \hline
Male & \thead{James, John, Robert, Michael, William, David,\\Richard, Joseph} \\ 
 \hline
Female & \thead{Mary, Patricia, Jennifer, Elizabeth, Linda,\\Barbara, Susan, Margaret} \\ 
 \hline
\end{tabular}
\caption{Items used for concepts and attributes}
\label{items}
\end{table}
} 

We chose \textit{computer science} and \textit{biology} as the two concepts, and \textit{male} and \textit{female} as the two attributes.  We selected these concepts and attributes because they reflect a well-known stereotype where it is reasonable to accept association between \textit{computer science} and \textit{male}.  Following the best practices described in \citet{Greenwald2005}, we selected words (see Table \ref{items}) to represent the concepts and attributes.

\subsection{Participants and Deception Strategies}

With the approval of our institutional review board\footnote{IRB\# 05-16-005}, we solicited participation by contacting friends and colleagues via direct email and private Facebook messaging.  Participants were directed to a webpage where they were presented with an informed consent agreement.  Those who agreed to participate were asked to complete a short demographic survey, which was not used in this specific study, followed by the IAT described above.  After completing the IAT (\textit{first attempt}), participants were shown their score and whether they associated \textit{computer science} with \textit{male} (positive score) or \textit{computer science} with \textit{female} (negative score).  Afterwards, participants were presented with an infographic that describes one of five randomly-selected deception strategies for altering one's score.

\begin{table}[!t]
\centering
 \begin{tabular}{|l|l|} 
 \hline
\textbf{\#} & \textbf{Description of deception strategy}\\
\hline
1 &  Make about 10 errors intentionally \\ 
 \hline
2 &  Say ``one Mississippi'' before pressing the appropriate key\\ 
 \hline
3 &  Put your hands in your lap between keypresses\\ 
 \hline
4 &  Cross your hands on the keyboard\\ 
 \hline
5 &  Touch your nose before pressing the appropriate key\\ 
 \hline

\end{tabular}
\caption{Five deception strategies}
\label{infographics}
\end{table}

\begin{figure}[!t]
\centering
\begin{subfigure}{.45\textwidth}
  \centering
  \includegraphics[width=.9\linewidth]{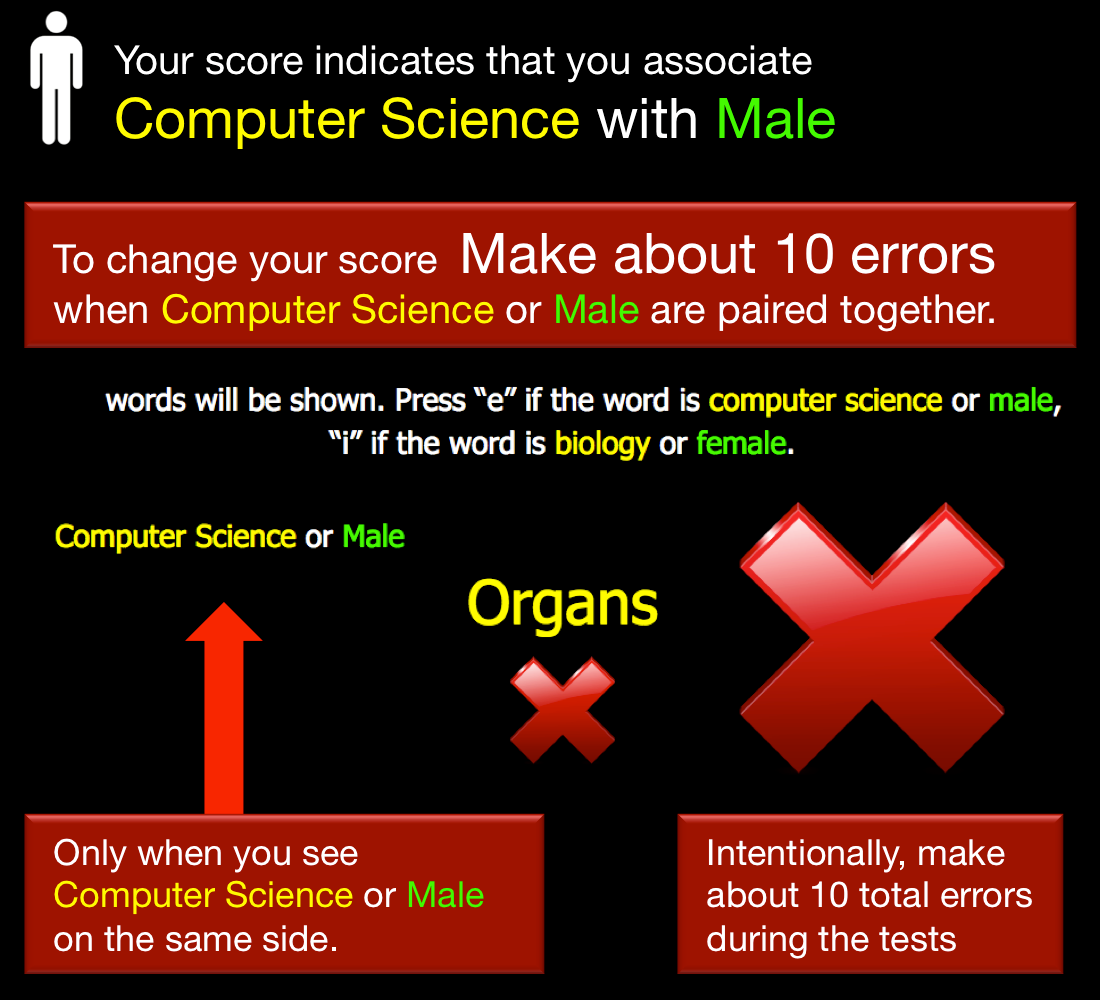}
  \caption{Instructions to disassociate computer science and male using error manipulation.}
\end{subfigure}%
\caption{Infographic Examples}
\label{infoexamples}
\end{figure}

Table \ref{infographics} summarizes the five deception strategies and Fig. \ref{infoexamples} shows an example infographic.  Each presented infographic is customized to describe when to use the deception strategy in order to alter one's score on the \textit{second attempt}.  For example, if a participant's \textit{first attempt} score is positive, the instructions (if followed) will yield a negative score. Conversely, if a participant's \textit{first attempt} score is negative, following the instructions will yield a positive score on the \textit{second attempt}.

\subsection{Method Goals}

Out methodology is designed to generate two  classes of IAT attempts: The \textit{first attempt} is where participants take our specific IAT for the first time and the \textit{second attempt}, is where participants have been previously shown (i) their \textit{first attempt} score, (ii) the association indicated by their score, and (iii) instructions on how to alter their score to achieve the opposite association.  Unlike previous studies \cite{Stieger2011,Agosta2010,Cvencek2010,Rohner2015}, we did not exclude \textit{first attempts} based on survey answers or direct observation nor did we exclude \textit{second attempts} that did not produce a significantly different score, i.e., unsuccessful faking/deception. We only removed attempts that were incomplete, i.e., the participant quit before finishing all the blocks in the two IAT attempts.  

While pruning the data based on direct observation and expert knowledge yields training data that is more accurate in terms of identifying natural implicit reaction vs. truly faked responses, consider that direct observation is challenging when administering online IATs and that participants might not be truthful when asked about previous IAT knowledge. 
Our goal is to show that previous methodologies may not be effective when applied to more realistic data and that a machine learning approach is needed to more accurately classify IAT results.

\section{IAT Attributes and Detecting Second Attempts}

An IAT records a participants key press response times and errors in categorizing $200$ items.  The items are presented in seven blocks where each block represents a screen configuration with different concepts, attributes, or both paired on the right and left side. Three of the blocks ($80$ presented items) are practice blocks that only include concepts or attributes to help familiarize the participant with the items and the correct classifications.  Four of the blocks ($120$ presented items) are critical blocks where both concepts and attributes are paired together on each side.  In critical blocks, the user's response time and accuracy are used to determine their IAT score.  In our experiment, all five of the presented deception strategies instruct the participant to use a delaying technique during the two critical blocks (60 presented items)
which paired the naturally associated concepts.
If the participant successfully employs that strategy, their score will be altered.  

Note that some participants may not employ the strategy at all, i.e., they may forget what to do on the second attempt.  In principle, these second attempts could be considered natural first attempts since the participant is not altering their behavior.  However, participants' response time and accuracy may change on the second attempt. Participants are more familiar with the items so response time and error rate could decrease, but they are also more fatigued and it may be harder to concentrate. Nonetheless, it may be possible to detect second attempts from response differences regardless of whether or not the deception strategy is being used effectively.  Also, note that some participants may employ the strategy during the wrong block configurations.  If the strategy is employed during practice blocks, it will not impact their IAT score at all, but it will impact response time and accuracy.  If the strategy is employed during the wrong critical blocks, it would exaggerate the score, i.e, strengthen association rather than reverse it.  Again, it may be possible to detect second attempts even if the deception strategy is being improperly employed.

It is important to note that second attempts do not necessarily represent deception.  Instead, they represent a more general and diverse class of IATs that are potentially compromised.  Correctly classifying first attempts vs. second attempts is likely more challenging than classifying verified natural attempts vs. effectively faked attempts.

\section{Results and Discussion}

\subsection{Overview}

Out of approximately $200$ solicited participants, $108$ agreed to the informed consent and $67$ completed all trials in both their \textit{first} and \textit{second} IAT attempts.  Table \ref{iatsummary} shows the mean and standard deviation of the response times, error rates, and IAT scores for both the \textit{first} and \textit{second} IAT attempts among the $67$ full participants. The last row shows the p-value of a two-tailed dependent \textit{t}-test for paired samples.  These values indicate a strong statistical difference between the first and second attempts.

On the first attempt, the average IAT score was $0.395$ and $87\%$ ($58$ out of $67$) participants scored above zero indicating that our participants tend to associate computer science with male. The information presented to the participants after the first attempt had a significant impact on altering the second attempt.   The mean score decreased to $0.010$ and only $52\%$ ($35$ out of $67$) participants scored above zero indicating a significant change to neutral association.
Overall $51$ out of $67$ participants were able to alter their score
by at least $1$ standard deviation opposite their initial association.

\begin{table}[!t]
\centering
 \begin{tabular}{|c|c|c|c|} 
 \hline
\textbf{Attempt} & \textbf{Response Time} & \textbf{Error Rate} & \textbf{Score} \\
\hline
\textbf{First} & $0.802\,(0.113)$ & $0.069\,(0.054)$ & $0.395\,(0.373)$ \\ 
\hline
\textbf{Second} & $0.844\,(0.168)$ &$0.096\,(0.066)$ &$0.010\,(0.500)$\\ 
\hline
\textbf{p-value} &  $0.0075$ & $0.0006$ & $<0.0001$\\ 
\hline
\end{tabular}
\caption{Mean value and standard deviation for the critical blocks of first and second attempts; p-value indicating  
the significance of the difference between the first and second attempts}
\label{iatsummary}
\end{table}

\subsection{Comparing Multiple Machine Learning Methods}

IAT Score data as well as individual trial times were exported from the website; using the
IAT package\footnote{https://cran.r-project.org/web/packages/IAT/index.html}
and basic functions in R, multiple features were calculated for each of the seven blocks, including the percent of errors, percent of responses faster than $300\text{ms}$, five-number summary, and skewness.
Using these measures, we made
two subsets of the dataset: an unpruned set consisting of all
IAT attempts and a pruned set of all first attempts
and only second attempts which reversed the score from the
first attempt.
We defined a successfully reversed score as a second IAT attempt that changed by at least one standard deviation toward the opposite direction of the initial attempt.
The final datasets did \textit{not} contain the score associated with
the IAT attempt as this would assume prior knowledge of the true score
of the attempt.

We performed feature selection on both subsets, and any attribute with a correlation value above $0.75$ was subsequently removed.  Removing the correlated attributes would allow for faster training times and eliminated any features that would not contribute to the models.  
For each subset, we used $7$ distinct machine learning algorithms provided by Weka (using the default parameters): naive Bayes, support vector machines, multinomial logistic regression, multilayer perceptron, simple logistic regression, propositional rule learner (JRip), and random forest.
Training and testing was done using leave-one-out cross-validation.
Table \ref{ml-summary} shows the weighted F1 scores across each subset and with each ML method.
We used F1 scores as a comparative metric because it is more
robust in comparison with accuracy, especially when looking
at datasets where classes are not equally distributed.

\begin{table}[!t]
\centering
 \begin{tabular}{|l|c|c|} 
 \hline
\textbf{ML Method} &  \textbf{Unpruned  ($n=154$)} & \textbf{Pruned ($n=102$)} \\ 
\hline
Naive Bayes &  0.721 & 0.735 \\ \hline
SVM &  0.728 & 0.779 \\ \hline
Logistic &  0.747 & 0.716 \\ \hline
Multilayer Perceptron &  0.719 & 0.812 \\ \hline
Simple Logistic &  0.700 & 0.764 \\ \hline
JRip &  0.675 & 0.712 \\ \hline
Random Forest &  0.678 & 0.745 \\ \hline
\end{tabular}
\caption{Results of various machine learning methods. Pruning the data entailed of removing deception attempts that resulted in a $<1$ SD score change.}
\label{ml-summary}
\end{table}


On the unpruned data, multinomial logistic performed the best
achieving an F1 score of $0.75$ followed closely by naive bayes,
SVM, and multilayer perceptron.
The multilayer perceptron performed the best on the pruned data achieving
an F1 score of $0.81$.
All models, with the exception of multinomial logistic, performed
better on the pruned data.
Overall, the more complex models (i.e., SVM and multilayer perceptron)
performed the best across the pruned and unpruned data.

It is important to note the model performances on both
the pruned and unpruned data since they address similar but
distinct tasks: the unpruned data looks to identify
unnatural or compromised IATs (ones that have not been taken
honestly and for the first time)
while the pruned data seeks to identify only
those who successfully reverse their score. Which one of these is relevant
would depend on the intended use of the model.

\subsection{Deeper Investigation with Muti-Layer Perceptrons}


In order to more closely investigate whether there were
more subtle complexities, we used a multilayer perceptron (MLP)
built in TensorFlow to more finely control the neural network
in order to see if offered any significant gains in F1 score
over the other methods.

The MLPs were tested with $10$-fold as well as leave-one-out cross-validation with
similar results.
After approximately $50$ epochs of training, the MLP would show significantly
lower training costs than on unseen test cases due to overfitting on the
training data; despite this, increases in performance occurred until about
$200$ epochs.
Even with methods of
regularization including reducing the hidden layer nodes,
adding dropout, and including weight decay in the cost metric failed
to significantly reduce overfitting in the neural net.
One likely cause of overfitting is the small training set size.
We found adding a second hidden layer did not alter the performance
of the MLP and would only worsen overfitting.
Normalizing the input datasets significantly improved the performance of the MLP.

Using a perceptron with one hidden layer written in TensorFlow
specifically with $13$ hidden nodes and a
$0.7$ keep probability (for dropout) attained an F1 score of
$0.78$ on the unpruned data after $200$ epochs (using a $0.5$ threshold for the F1 score).
This F1 score outperformed all of the Weka models.
The TensorFlow MLP
was able to match but not exceed the Weka MLP with an F1 score
of $0.81$ on the pruned data.
Although pruning improved the performance of the model,
the MLP does not rely on pruning for effective prediction.
While we specifically compared F1 scores, accuracies for the
TensorFlow MLP fell in the $0.7$ to $0.8$ range.

\subsection{Comparison to Previous Methods of Detection}

As a point of comparison, we implemented the top-performing
method of faking detection presented in \citet{Agosta2010} which
is based off of the ratio of response times of the fastest pair
of critical blocks to those of the corresponding practice
blocks. Table \ref{agosta-table} shows our top performing model
(the TensorFlow MLP) compared against the ratio-based model both
on the unpruned and pruned data. Tweaking the threshold ratio
for faked/non-faked IATs did not significantly alter the F1 score.

The ratio-based model showed accuracies of $70\%$ or lower as opposed to $80\%$ as presented in \citet{Agosta2010} suggesting that it might not be robust against different deception strategies. Our machine learning models demonstrate better robustness
to different deception methods
by significantly outperforming the ratio-based model.

\begin{table}[!t]
\centering
\begin{tabular}{|l|c|c|}
\hline
\textbf{Model} & \textbf{F1 Unpruned} & \textbf{F1 Pruned} \\ \hline 
MLP & $0.78$ & $0.81$ \\ \hline
Ratio & $0.64$ & $0.58$ \\
\hline
\end{tabular}
\caption{F1 scores for TensorFlow MLP and ratio-based method from \cite{Agosta2010}}
\label{agosta-table}
\end{table}

\section{Conclusion}



Simple data analysis revealed that $51$ out of $67$ participants were
able to significantly alter their score with only brief training on how to do so.
This clearly motivates the need for a way to detect compromised IATs as they
are not difficult to deceive.
	
We found that more complex models such as SVM and MLP perform
slightly better the simpler models we tested. This suggests that there
might be some nuances in the data that require a more sophisticated model
to detect. Yet the fact that an MLP overfits quickly on the data even with
regularization suggests that more data would be needed to make full use
of the model.

With F1 scores (along with accuracy, precision, and recall) in the $0.7$ to $0.8$
range, the models we trained would not prove accurate enough to determine if
a single IAT attempt was honest or not (e.g., $1$ in $5$ attempts is a false positive
or a false negative). Yet the models could still provide an accurate assessment
of the quality of a large set of IAT attempts, which would still prove useful
in research applications in psychology and cognitive science.
Furthermore, while our models provide comparable performance
to previous attempts to identify faked IATs, using machine
learning to do so offers a more robust solution regardless of the use of pruning or the method of deception.

Future work in this area of research could address two main issues.
First, it could determine whether more data is able to improve the
performance of more complex machine learning models.
Second, it could address whether a model trained on data from one
IAT could be applied directly to another IAT or if each IAT must
have a model trained only on that IAT.

Overall, the task of detecting unnaturally taken IATs is well suited for machine learning as it is a well defined classification problem supported by a rich array of data points and aggregate metrics. That being said, the nature of obtaining data directly from human subjects can present two challenges.
First, machine learning models perform better with a large
volume of data, yet obtaining data from human subjects is 
time-consuming and requires a relatively large amount of effort. Second, the veracity of the data will always be in question; this is exaggerated by the fact that IATs rely very heavily on \textit{honest, natural} responses which can never be undisputedly verified.
Despite this, we have developed a method of identifying compromised
IATs that is robust in the face of multiple different deception strategies.

\bibliographystyle{plainnat}
\bibliography{main}

\begin{thebibliography}{17}
\providecommand{\natexlab}[1]{#1}
\providecommand{\url}[1]{\texttt{#1}}
\expandafter\ifx\csname urlstyle\endcsname\relax
  \providecommand{\doi}[1]{doi: #1}\else
  \providecommand{\doi}{doi: \begingroup \urlstyle{rm}\Url}\fi

\bibitem[Agosta et~al.(2010)Agosta, Ghirardi, Zogmaister, Castiello, and
  Sartori]{Agosta2010}
S.~Agosta, V.~Ghirardi, C.~Zogmaister, U.~Castiello, and G.~Sartori.
\newblock {Detecting Fakers of the autobiographical IAT}.
\newblock \emph{{Applied Cognitive Psychology}}, 25:\penalty0 1299--1306, 2010.

\bibitem[Bargh(2013)]{Bargh2013}
John~A. Bargh.
\newblock \emph{{Social psychology and the unconscious: The automaticity of
  higher mental processes}}.
\newblock Psychology Press, 2013.

\bibitem[by~Digital~Science(2018)]{Dimensions}
Dimensions by~Digital~Science.
\newblock Publication per year for ``implicit association test''.
\newblock 2018.
\newblock URL
  \url{https://app.dimensions.ai/analytics/publication/viz/overview-publications?search_text=implicit\%20association\%20test}.

\bibitem[Cvencek et~al.(2010)Cvencek, Greenwald, Brown, Gray, and
  Snowden]{Cvencek2010}
D.~Cvencek, A.~Greenwald, A.~Brown, N.~Gray, and R~Snowden.
\newblock {Faking of the Implicit Association Test Is Statistically Detectable
  and Partly Correctable}.
\newblock \emph{{Basic and Applied Social Psychology}}, 32:\penalty0 302--314,
  2010.

\bibitem[Fiedler and Bluemke(2010)]{Fiedler2010}
K.~Fiedler and M.~Bluemke.
\newblock {Faking the IAT: Aided and Unaided Response Control on the Implicit
  Association Tests}.
\newblock \emph{{Basic and Applied Social Psychology}}, 27:\penalty0 307--316,
  2010.

\bibitem[Greenwald and Banaji(1995)]{Greenwald1995}
A.~G. Greenwald and M.~R. Banaji.
\newblock {Implicit social cognition: Attitudes, self-esteem, and stereotypes}.
\newblock \emph{{Psychological Review}}, 102:\penalty0 4--27, 1995.

\bibitem[Greenwald et~al.(1998)Greenwald, McGhee, and Schwartz]{Greenwald1998}
A.~G. Greenwald, D.~E. McGhee, and J.~L. Schwartz.
\newblock {Measuring individual differences in implicit cognition: the implicit
  association test}.
\newblock \emph{{Journal of Personality and Social Psychology}}, 74:\penalty0
  1464--1480, 1998.

\bibitem[Greenwald et~al.(2003)Greenwald, Banaji, and Nosek]{Greenwald2003}
A.~G. Greenwald, M.~R. Banaji, and B.~A. Nosek.
\newblock {Understanding and using the Implicit Association Test: I. An
  improved scoring algorithm}.
\newblock \emph{{Journal of Personality and Social Psychology}}, 85:\penalty0
  197--216, 2003.

\bibitem[Greenwald et~al.(2005)Greenwald, Banaji, and Nosek]{Greenwald2005}
A.~G. Greenwald, M.~R. Banaji, and B.~A. Nosek.
\newblock {Understanding and using the Implicit Association Test: II. Method
  variables and construct validity}.
\newblock \emph{{Journal of Personality and Social Psychology}}, 31:\penalty0
  166--180, 2005.

\bibitem[Greenwald et~al.(2009)Greenwald, Poehlman, Uhlmann, and
  Banaji]{Greenwald2009}
A.~G. Greenwald, T.~A. Poehlman, E.~L. Uhlmann, and M.~R. Banaji.
\newblock {Understanding and using the Implicit Association Test: III.
  Meta-analysis of predictive validity}.
\newblock \emph{{Personality and Social Psychology Bulletin}}, 97:\penalty0
  17--41, 2009.

\bibitem[Hofmann et~al.(2005)Hofmann, Gawronski, Gschwendner, Le, and
  Schmitt]{Hofmann2005}
W.~Hofmann, B.~Gawronski, T.~Gschwendner, H.~Le, and M.~Schmitt.
\newblock {A Meta-Analysis on the Correlation Between the Implicit Association
  Test and Explicit Self-Report Measures}.
\newblock \emph{{Personality and Social Psychology Bulletin}}, 31:\penalty0
  1369--1385, 2005.

\bibitem[McDaniel et~al.(2009)McDaniel, Beier, Perkins, Goggin, and
  Frankel]{McDaniel2009}
M.~McDaniel, M.~Beier, A.~Perkins, S.~Goggin, and B.~Frankel.
\newblock {An assessment of the fakeability of self-report and implicit
  personality measures}.
\newblock \emph{{Journal of Research in Personality}}, 43:\penalty0 682--685,
  2009.

\bibitem[Nosek et~al.(2011)Nosek, Beth~Hawkins, and S~Frazier]{Nosek2011}
Brian Nosek, Carlee Beth~Hawkins, and Rebecca S~Frazier.
\newblock Implicit social cognition: From measures to mechanisms.
\newblock \emph{Trends in cognitive sciences}, 15:\penalty0 152--9, 03 2011.

\bibitem[R\"{o}hner and Torsten(2015)]{Rohner2015}
J.~R\"{o}hner and E.~Torsten.
\newblock {Trying to separate the wheat from the chaff: Construct- and
  faking-related variance on the Implicit Association Test (IAT)}.
\newblock \emph{{Behavior Research Methods}}, 48:\penalty0 243--258, 2015.

\bibitem[R\"{o}hner et~al.(2013)R\"{o}hner, Schr\"{o}der-Ab\'{e}, and
  Sch\"{u}tz]{Rohner2013}
J.~R\"{o}hner, M.~Schr\"{o}der-Ab\'{e}, and A.~Sch\"{u}tz.
\newblock {What do fakers actually do to fake the IAT? An investigation of
  faking strategies under different faking conditions}.
\newblock \emph{{Journal of Research in Personality}}, 47:\penalty0 330--338,
  2013.

\bibitem[Steffens(2004)]{Steffens2004}
M.~Steffens.
\newblock {Is the Implicit Association Test Immune to Faking?}
\newblock \emph{{Experimental Psychology}}, 51:\penalty0 165--179, 2004.

\bibitem[Stieger et~al.(2011)Stieger, G\"{o}ritz, Hergovich, and
  Voracek]{Stieger2011}
S.~Stieger, A.~G\"{o}ritz, A.~Hergovich, and M.~Voracek.
\newblock {Intentional faking of the single category Implicit Association Test
  and the Implicit Association Test}.
\newblock \emph{{Psychological Reports}}, 109:\penalty0 219--230, 2011.

\end{thebibliography}

\end{document}